# Identifying Bridges from Asymmetric Load-Bearing Structures in Tapped Granular Packings

Chijin Zhou,[1] Shuyang Zhang,[2] Xueliang Dai,[1] Yixin Cao,[2] Ye Yuan,[4] Chengjie Xia,[5] Zhikun Zeng,[2,*] and Yujie Wang[1,2,3,†]

[1]*Department of Physics, College of Mathematics and Physics, Chengdu University of Technology, Chengdu 610059, China*
[2]*School of Physics and Astronomy, Shanghai Jiao Tong University, Shanghai 200240, China*
[3]*State Key Laboratory of Geohazard Prevention and Geoenvironment Protection, Chengdu University of Technology, Chengdu 610059, China*
[4]*Research Center for Advanced Science and Technology, University of Tokyo, Tokyo 153-8505, Japan*
[5]*Shanghai Key Laboratory of Magnetic Resonance, School of Physics and Electronic Science, East China Normal University, Shanghai 200241, China*

Using high-resolution x-ray tomography, we experimentally investigate the bridge structures in tapped granular packings composed of particles with varying friction coefficients. We find that gravity can induce subtle structural changes on the load-bearing contacts, allowing us to identify the correct load-bearing contacts based on structural information alone. Using these identified load-bearing contacts, we investigate the cooperative bridge structures which are mechanical backbones of the system. We characterize the geometric properties of these bridges and find that their cooperativity increases as the packing fraction decreases. The knowledge of bridges can enhance our understanding of the rheological properties of granular materials.

Granular packings are ubiquitous in nature and have diverse industrial applications. Understanding their load-bearing properties is crucial, as these properties significantly affect the stability and rheological properties of granular materials under gravity or other external loads [1]. When subjected to external forces, the contact network in granular systems reorganizes to form heterogeneous load-bearing structures, such as bridges (or arches). Bridges are collective structures in which neighboring grains rely on each other for mutual stability [2-6]. In real-world systems, the presence of gravity and friction leads to the formation of arch-like bridge structures. The bridge configuration enables the efficient distribution of external loads across all constituting particles. Additionally, due to the interlocking nature of neighboring particles in a bridge, bridges are rigid and capable of withstanding imposed loads [7,8], a fact known since ancient times. The presence of bridges can fundamentally alter the material's mechanical and rheological properties. For instance, bridge formation within or at the outlet of a hopper can induce clogging during granular flow, a phenomenon commonly encountered in the handling of granular materials [9-11]. The fact that the clogging structures are always arch-like suggests that once formed, bridges behave more like solid structures and can resist more external loads compared to other structure configurations. Therefore, a comprehensive knowledge of bridges is indispensable for understanding the structural and rheological behaviors of frictional granular systems [12].

Random granular packings also display force heterogeneity [13], manifested as the existence of spatially inhomogeneous force chain networks that have been extensively investigated [14,15]. The force chain networks consist of both strong and weak force chains, with the strong force chains traditionally regarded as the mechanical backbones of granular packings. Nevertheless, recent work has shown that there exists no strong correlation between strong force chains and local rigidity, as identified by pebble game and dynamical matrix analysis [7,8]. Instead, bridges are frequently observed at the boundaries of rigid clusters [7]. These findings challenge the long-standing assumption that bridges and strong force chains are closely related or are even manifestations of the same rigid network within the system. Moreover, unlike bridges, strong force chain structures are not always arch-like, which suggests they might not always possess the capacity to resist external loads, leaving the relationship between these two even more obscure and implying that there may not be a one-to-one correspondence between them [16]. While much of the previous work has focused on force chains, almost treating strong force chains as synonymous with the mechanical backbones of granular materials, the study of bridges has received insufficient attention despite its significance.

Previously, studies on bridge structures in granular packings have been mainly limited to numerical simulations [5,6], with few experimental investigations due to the challenges in determining contact networks in three-dimensional (3D) granular packings [17]. Notably, bridges have also been identified in dense particulate packings like colloids [18,19], and it has been shown that the structures of bridges identified are quite insensitive to gravity or mechanical load, implying that these structures are universal in dense packings and that applied load only exploits the existing contact network to generate mechanical stability [18]. However, one limitation of the work is the rather poor experimental spatial resolution exploited. Consequently, the load-bearing structures are selected by empirical assumption rather than direct experimental evidence. Given that there exist many alternative ways to define the load-bearing structures, it remains unclear why this specific criterion was chosen. Without a reliable method for identifying bridge structures, the key conclusions become less convincing.

In this article, we use X-ray tomography to obtain high-spatial-resolution 3D structures of tapped granular packings composed of particles with different friction coefficients. We identify a clear up-down structural asymmetry at the particle contact level, induced by load-bearing contacts that resist gravitational forces. Using this asymmetry, we are able to choose the appropriate criterion, the LCOM method, to identify the load-bearing contacts in the system based on structural information alone. Subsequently, we characterize the geometric properties of cooperative bridge structures and reveal an increase in cooperativity as the volume fraction decreases. The knowledge of bridges can greatly enhance our understanding of the rheological properties of granular materials.

We utilize three types of monodisperse beads to prepare disordered granular packings through mechanical tapping. These beads, having different surface friction coefficients, are acrylonitrile butadiene styrene (ABS) plastic beads ($\mu = 0.52$), 3D-printed (3DP) plastic beads ($\mu = 0.66$), and 3D-printed plastic beads with a bumpy surface (BUMP) ($\mu = 0.86$). The diameters of ABS and BUMP beads are $d = 6$ mm, with 3DP bead sizes of $d = 5$ mm. Packings are prepared in a cylindrical container with an inner diameter of 140 mm. The bottom and side walls of the container are roughened by randomly gluing ABS hemispheres to prevent crystallization. The packing height is approximately 200 mm. We tap the initial packings using a mechanical shaker with different tapping intensities to generate a full range of packing fractions between random loose packing (RLP) and random close packing (RCP). See Ref. [20]

for further details on the tapping methods. Subsequently, we acquire the three-dimensional (3D) packing structures using a medical CT scanner (UEG Medical Group Ltd., 0.2 mm spatial resolution). The coordinates and size of each particle are determined with an error of less than $3\times10^{-3}d$ using well-developed image processing techniques similar to our previous studies [21]. The high spatial resolution ensures accurate detection of particle contacts, which is critical for the current study (see Supplemental Material [22] for more details). After discarding particles located within $2.5d$ from the container boundary, a total of 3000~6000 particles are included in the following analysis. We repeat each experiment 10~30 times at each tapping intensity to enhance statistical reliability.

Once the contact network is established, the next critical step is identifying the force-bearing neighbors among all contact ones. In a mechanically stable packing under gravity, a particle is generally considered to be supported by a base of three contact neighbors, with the requirement that the projection of the particle's center-of-mass falls within the triangle formed by these three base particles [Fig.1(a)]. However, the challenge in accurately identifying the true mechanical support base using this simple criterion lies in the fact there exist multiple possible combinations of three particles satisfying the above support-base requirements, with no perfect method currently available to select the correct one. If, as suggested by the colloidal experiments, gravity only exploits the existing contact network to generate load-bearing structures without modifying them [18,19], it would, in principle, be impossible to identify the correct support base through structural analysis alone. However, if the force-bearing contacts differ from other contacts in a certain way due to the influence of gravity, it is therefore possible to identify them by searching for gravity-induced footprints.

To achieve this, we analyze the packing structures in various orientations relative to gravity, namely, the real gravity direction (0°), four orthogonal directions within the horizontal plane (90°), and the reverse gravity direction (180°), to see whether there exist noticeable differences. We first calculate the average number of possible supporting bases $N_{base}$ for each particle in different packings. Note that particles that have at least four contact neighbors and at least one support base are analyzed. As shown in Fig. 1(b), $N_{base}$ increases monotonically with the average contact number $Z$, which can be simply explained by the fact that more contact neighbors provide more possible choices of support bases. Notably, $N_{base}$ determined in the real gravity direction (0°) is about 0.5 less than that obtained from the reverse gravity direction (180°), suggesting the existence of certain up-down asymmetry at the particle contact level. We emphasize that this asymmetry is beyond experimental uncertainty, as the values obtained from the four orthogonal horizontal directions (90°) are almost identical and lie exactly between the values for 0° and 180°. The observed up-down asymmetry implies that gravity can noticeably influence the positions of the load-bearing contacts, and it is therefore possible to identify them based on their differences from non-load-bearing ones. Nevertheless, the information conveyed by $N_{base}$ is not straightforward, as it does not imply that each particle has more contacting neighbors above than below [25].

To understand how true load-bearing bases contribute to the observed up-down asymmetry, we calculate the solid angle $\Omega$ formed by the vectors connecting a center particle to its three contacting neighbors in a support base. Figure 2(a) shows the probability distribution functions (PDFs) of $\Omega$ for all possible bases $P(\Omega)$ for real gravity and other imaginary gravity directions. These distributions generally exhibit a peak at $\Omega \sim \pi/4$, followed by a hump around $\Omega = \pi/2$ and a long tail extending to larger solid angles. This shape originates from the full contact network where most contacts are not force-bearing. Nevertheless, subtle differences among distributions for real and imaginary gravity directions are evident, where more weight is observed at the hump around $\Omega = \pi/2$ in $P_{0°}(\Omega)$ (real gravity direction) compared to the other two distributions. It is natural to presume that this difference $\Delta P(\Omega) = P_{0°}(\Omega) - P_{90°}(\Omega)$ originates from gravity-bearing structures and that $\Delta P$ should be related to the proportion of load-bearing contacts among all contacts. Specifically, gravity seems to move load-bearing bases with initial $\Omega \sim \frac{\pi}{4}$ to $\Omega \sim \frac{\pi}{2}$ [inset of Fig. 2(a)]. Furthermore, we integrate $\Delta P(\Omega)$, $\omega = \int_0^{2\pi} |\Delta P/2| \, d\Omega$, to quantify the proportion of gravity-induced load-bearing contacts relative to all contacts. As shown in Fig. 2(d), $\omega$ decreases as $Z$ increases, a trend aligning with the expectation that looser packings require a higher proportion of force-bearing contacts to maintain mechanical stability.

With information gained from the above analysis, it is possible for us to choose the correct force-bearing bases from all potential ones. Several criteria based solely on structural information have been previously proposed. One method usually adopted in simulation studies is the "lowest center of mass (LCOM)" method. This method selects the support base as the one possessing the lowest average centroid among all possible bases [6,26]. The validity of LCOM is straightforward in the case of packings prepared by sequential deposition, where each particle has to sit on top of the lowest three contact neighbors to maintain mechanical stability. However, its applicability to tapped packings is not justified *a prior*. In these systems, all the particles come to rest almost simultaneously after each tap, and it is uncertain whether the load-bearing base is still the one possessing the lowest center of mass. In colloidal experiments, an alternative "lowest mean-squared separation (LSQS)" method has been introduced [18,19]. The LSQS selects the support base with the smallest mean-squared separation distance from the supported particle. This method was introduced to minimize the influence of many mistakenly identified contacts due to the limited spatial resolution in colloidal experiments. In the following, we use both LCOM and LSQS to identify load-bearing bases and determine which one is more compatible with the experimentally observed up-down asymmetry. Additionally, we also randomly select a base among all possible ones as the load-bearing one (RANDOM) for comparison.

The PDFs of $\Omega$ for bases selected using the LOCM method $P_{\text{LCOM}}(\Omega)$ are shown in Fig. 2(b). We observe that $P_{\text{LCOM}}(\Omega)$ for 0° bears a great resemblance to the experimentally obtained $\Delta P(\Omega)$, both featuring a peak at $\Omega_p \sim \pi/2$ [inset of Fig. 2(b)], along with a tail at larger $\Omega$. Additionally, we note a clear asymmetry in $P_{\text{LCOM}}(\Omega)$ between the gravity and reverse-gravity directions. In contrast, both the LSQS and RANDOM methods exhibit almost negligible up-down asymmetry [Fig. 2(c)], indicating that they have misidentified many non-load-bearing bases as load-bearing ones. The rather similar behavior between LSQS and RANDOM also suggests that LSQS essentially selects base randomly. Therefore, LCOM is the more appropriate method for identifying load-bearing structures as it is more capable of revealing gravity-induced asymmetry.

After identifying the force-bearing neighbors using LCOM, we further analyze bridges, *i.e.*, collective force-bearing structures in our system. Bridge structures consist of particles that are mutually supportive, meaning two contact particles are part of each other's support base [Fig. 1(a)]. We first examine the fraction of particles $f_b$ that are part of bridges. For LCOM, $f_b$ decreases with increasing $Z$ and is proportional to $\omega$ [Fig. 2(d)]. This suggests that bridges are directly responsible for the observed up-down asymmetry and are more abundant in low-$\phi$ packings. As $\phi$ increases, bridges collapse and form configurations where particles are no longer mutually supportive [6]. To directly verify the correlation between bridges and up-down asymmetry, we also prepared granular packings by the raining method, where particles are added sequentially. Further details on the raining experiments can be found in Ref. [24]. A significantly reduced up-down asymmetry is observed at similar $\phi$ compared with tapped packings [Fig. 2(d)]. In comparison, for LSQS and RANDOM, $f_b$ increases with $Z$ [inset of Fig. 2(d)], leading to the unphysical conclusion that large-$\phi$ packings contain more bridge particles. This contradicts both the trend observed in the asymmetry analysis and physics intuition. In retrospect, when LSQS was employed to analyze bridge structures in colloidal experiments [18,19], it was concluded that gravity plays a negligible role in bridge formation. This discrepancy most likely arises from the subtle influence of gravity on load-bearing structures, which is beyond the experimental resolution of the colloidal experiments. Without sufficient resolution, ambiguity in selecting the correct support base makes the subsequent bridge analysis challenging.

Next, we investigate the structural cooperativity associated with bridges by analyzing the bridge length distributions in different packings. Bridges can be either linear or complex depending on the presence of loops and/or branches [6]. Since most bridges in our packings are short and generally linear in shape, we focus on linear bridges. Assuming that the probability for a linear bridge to extend from length $n$ to $n+1$ is $p$, the corresponding length distribution follows $P_{\text{lin}}(n) = \alpha \exp(-\alpha n)$, in which $p = \exp(-\alpha)$. A larger $p$ value suggests greater cooperativity in the system [27]. We fit the PDFs of linear bridge lengths calculated using LCOM, as shown in Fig. 3(a). For LCOM, $p$ and the average bridge length $\langle n \rangle$ increases as $Z$ decreases, suggesting enhanced cooperativity in looser packings [Fig. 3(b)]. We also present results using LSQS and RANDOM, which show an almost constant $p$ dependency on $Z$. It is worth noting that both $f_b$ and $p$ are solely dependent on $Z$, irrespective of friction coefficient $\mu$ or packing fraction $\phi$. This remains true even when LSQS or RANDOM are used, though with different values of $f_b$ and $p$. This result is consistent with previous observations that the topological properties of the contact network, are primarily governed by $Z$ of the system. Bridges only exploit the contact network in specific ways and still have to conform to the inherent properties of the network [18].

To further analyze the geometric features of bridge structures, we examine the angle $\theta$ formed by the contact vector connecting bridge particles to their support base particles, relative to the gravity direction. Figure 4(a) shows the distributions of $\theta$, $P(\theta)$, for base particles selected by the three methods. For LCOM, $\theta$ is more concentrated near the vertical direction compared to LSQS and RANDOM, which is expected since LCOM tends to select support particles with a lower center of mass. Given that particles in bridges are mutually supportive, $\theta$ for consecutive particles in a bridge should satisfy $\theta_{i,i+1} + \theta_{i+1,i} = \pi$, and the mean extension probability of a linear bridge can then be estimated by $p = 2\int_0^\pi P(\theta)P(\pi-\theta)d\theta$ [curves in Fig. 3(d)], which is consistent with $p$ calculated from the bridge length distributions. Additionally, as shown in Fig. 4(b), $\langle \theta_i \rangle$, the average value of $\theta$ for the $i$-th particle in bridges with a given bridge length, decreases as $i$ increases. This trend suggests that bridges generally adopt a globally arch-like shape [9], a configuration well-known for its mechanical stability against gravitational loads. To quantify the degree of archness of bridges, we define the archness parameter $\alpha = \dfrac{h}{d}$, where $h$ and $d$ are the vertical and horizontal projected lengths of a bridge [Fig. 4(c)]. As $Z$ or $\phi$ decreases, $\alpha$ increases noticeably, indicating that bridges become more arched. Notably, there exists a universal correspondence between $\alpha$ and $\phi$ for all packings, rather than with $Z$, which suggests that $\alpha$ is primarily determined by the particle packing structure instead of the contact topology. This conclusion is further supported by the observation that there exists almost no difference between the local $\phi$ of bridge particles and that of non-bridge particles [Fig. 4(d)].

In summary, using 3D packing structures obtained from high-spatial-resolution X-ray tomography, we investigate the bridge structures in tapped granular packings. We clearly observe that gravity can induce a strong imprint on the load-bearing contacts, which helps us identify them based on structural analysis alone. We further characterize the geometric properties of the collective bridge structures, which are the mechanical backbones of the system. Since their omnipresence in granular materials and generally greater rigidity than the background, bridges are expected to contribute significantly to the complex rheological behaviors of granular materials [1,28].

*Acknowledgments*—The work is supported by the National Natural Science Foundation of China (No. 12274292, No. 123B2060).

*Corresponding author.

zzk97115_kenny@sjtu.edu.cn

†Corresponding author.

yujiewang@sjtu.edu.cn

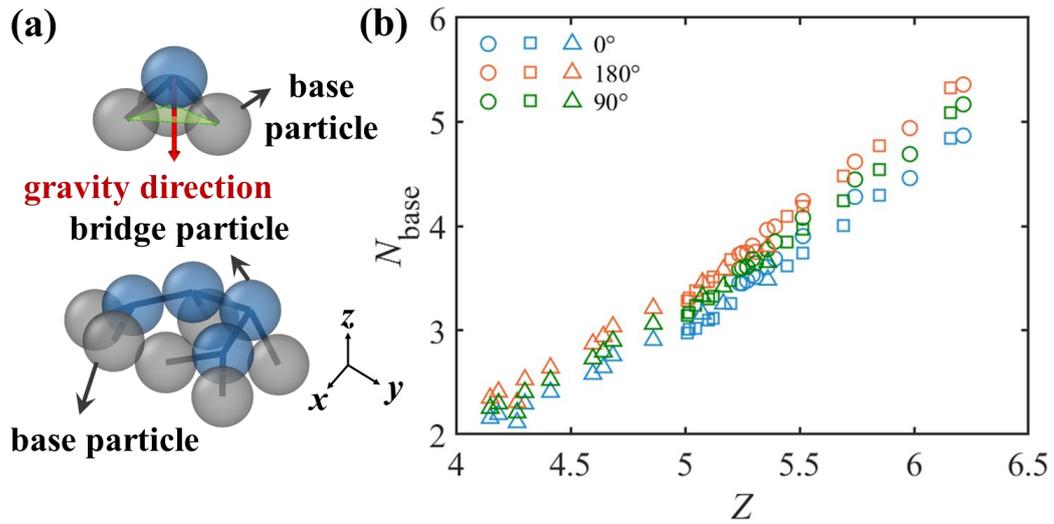

FIG. 1. (a) Schematic diagram of the support base (upper) and bridge structure (lower). (b) Average number of possible supporting bases $N_{base}$ for each particle as a function of the average contact number $Z$. Different symbols denote calculations for real gravity direction (0°), reverse gravity direction (180°), and orthogonal directions in the horizontal plane (90°) in packings composed of ABS (circles), 3DP (squares), and BUMP (triangles) grains.

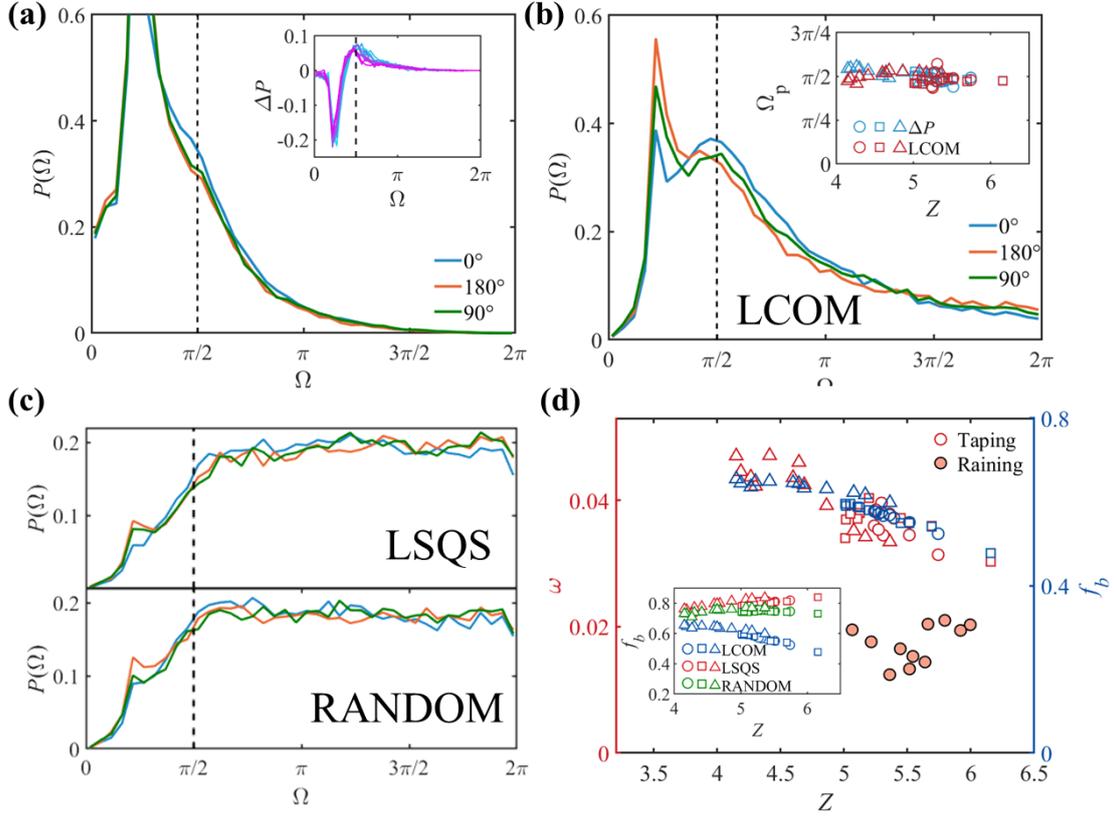

FIG. 2. (a) PDFs of $\Omega$ for all possible support bases for 0°, 180°, and 90°. Inset: the weight differences $\Delta P(\Omega) = P_{0°}(\Omega) - P_{90°}(\Omega)$ for BUMP systems with $Z$ ranging from 4.15 to 5.35 (from red to blue). The dotted lines are guides to the eye. (b, c) PDFs of $\Omega$ for bases selected using the LOCM (b), LSQS [upper in (c)], and RANDOM [lower in (c)] methods for different gravity directions. The dotted lines are guides to the eye for the hump at $\Omega \sim \pi/2$. Results are for BUMP systems with $Z = 5.35$. Inset of (b): locations of humps in $\Delta P(\Omega)$ and $P_{\text{LCOM}}(\Omega)$ for different packings. (d) Proportion of gravity-induced load-bearing contacts among all contacts $\omega$ (red, left axis) and the fraction of bridge particles $f_b$ obtained using LCOM (blue, right axis) for different packings. The solid symbols are for packings prepared by the raining method. Inset: $f_b$ for the LCOM, LSQS, and RANDOM methods as functions of $Z$.

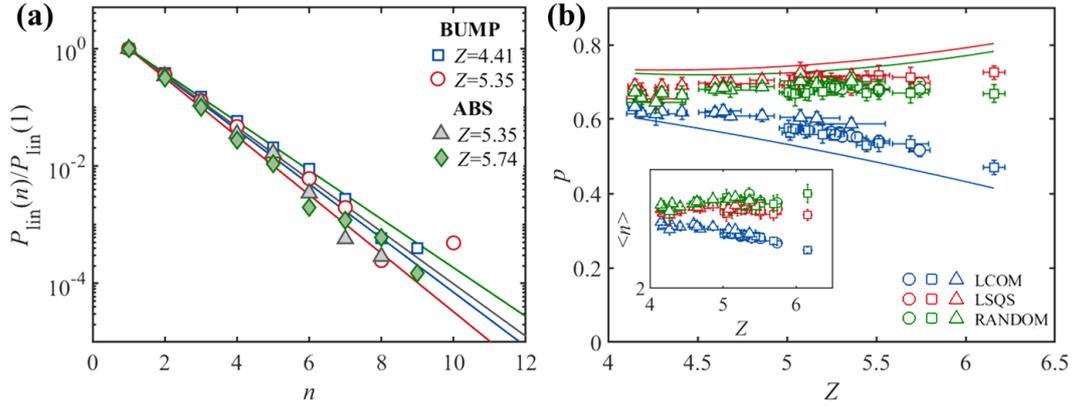

FIG. 3. (a) Length distributions of linear bridges $P_{\text{lin}}(n)$ identified using the LCOM method for different packings. The solid lines are fittings using $P_{\text{lin}}(n) = \alpha \exp(\alpha n)$. (b) Extension probability $p$ and average bridge length $\langle n \rangle$ (inset) for the LCOM, LSQS, and RANDOM methods in different packings. The solid curves correspond to the results of the theoretical model.

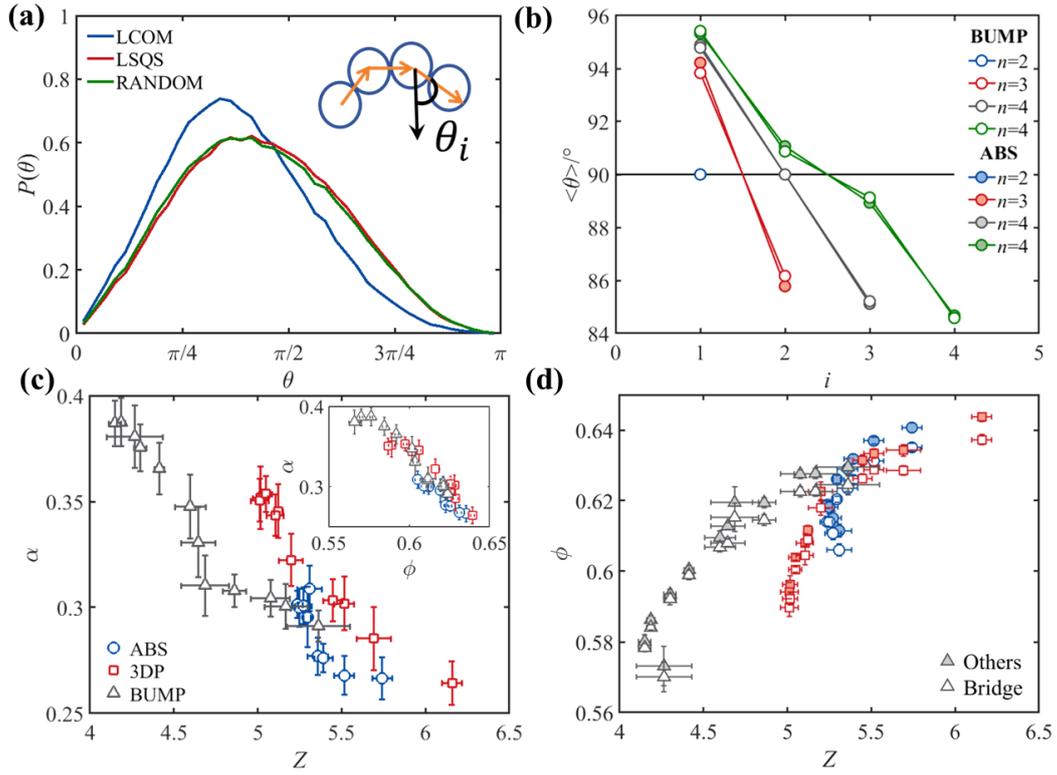

FIG. 4. (a) PDFs of $\theta$ for the LCOM, LSQS, and RANDOM methods. Inset: schematic diagram of a linear bridge structure. (b) Average angle $\langle\theta_i\rangle$ for bridges of a given length in different packings. The solid line is a guide to the eye for $\langle\theta_i\rangle=90°$. (c) Degree of archness $\alpha$ for bridges as a function of $Z$ and of $\phi$ (inset) for different packings. (d) Local volume fraction $\phi$ of bridge and other particles as a function of $Z$.

# Supplemental Materials for Identifying Bridges from Asymmetric Load-Bearing Structures in Tapped Granular Packings

Chijin Zhou,[1] Shuyang Zhang,[2] Xueliang Dai,[1] Yixin Cao,[2] Ye Yuan,[4] Chengjie Xia,[5] Zhikun Zeng,[2,*] and Yujie Wang[1,2,3,†]

[1]*Department of Physics, College of Mathematics and Physics, Chengdu University of Technology, Chengdu 610059, China*
[2]*School of Physics and Astronomy, Shanghai Jiao Tong University, Shanghai 200240, China*
[3]*State Key Laboratory of Geohazard Prevention and Geoenvironment Protection, Chengdu University of Technology, Chengdu 610059, China*
[4]*Research Center for Advanced Science and Technology, University of Tokyo, Tokyo 153-8505, Japan*
[5]*Shanghai Key Laboratory of Magnetic Resonance, School of Physics and Electronic Science, East China Normal University, Shanghai 200241, China*

*Corresponding author.

zzk97115_kenny@sjtu.edu.cn

†Corresponding author.

yujiewang@sjtu.edu.cn

## 1. Determination of particle contact

Packing structures obtained from the CT scans are used to determine inter-particle contact information. The high spatial resolution ensures accurate detection of particle contacts, which is critical for the current study. Ideally, the distance between the surfaces of two particles should be zero in actual contact. However, a variety of experimental uncertainties caused by factors such as the limited X-ray spatial resolution, particle non-sphericity, and artifacts in the image processing can result in them being inaccurately identified as having gaps or even penetrating each other. To overcome this issue, a standard methodology involving a complementary error function fitting program is followed to determine particle contacts within the system [23]. The PDF of surface-to-surface distance features a Gaussian core (red curve) and a fat tail. The Gaussian core arises from experimental uncertainties regarding contacting particles, where two particles in contact may be inaccurately identified as either penetrating or having a gap from each other. The fat tail at $\Delta r/d > 0$ stems from the contribution of neighboring but non-contacting particles. By employing a certain threshold value, represented by $\delta$, we consider particles to be in contact if their surface-to-surface distance $\Delta r$ falls below this value. The relationship between average contact number $Z$ and the threshold value $\delta$, as shown by blue symbols, is the cumulative behavior of the PDF of surface-to-surface distance distribution in (a), which can be described by fitting it with an error function for experimental uncertainties on contacting particles (red curve) and a linear behavior for non-contacting neighbors (green curve) as shown by the blue symbols. By fitting the experimental distribution with the above model (yellow curve), we can obtain the correct value of $Z_c$ and then extract the $\delta_c$. Consequently, particles are considered to be in contact when their gap distance is less than $\delta_c$.

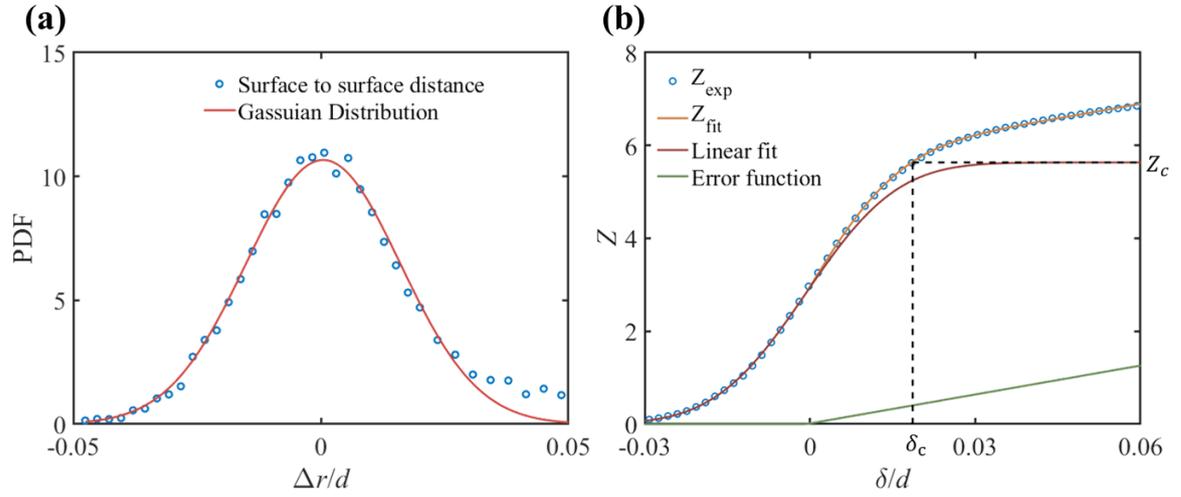

FIG. S1. (a) PDF of surface-to-surface distance $\Delta r$ among neighboring particles. The Gaussian core results from experimental uncertainties on contacting particles while the fat right tail stems from the neighboring but non-contacting particles. (b) The complementary error function fitting to yield the critical threshold $\delta_c$ of surface-to-surface distance and the corresponding global average contact number $Z_c$.